\documentclass[aip,jcp,reprint]{revtex4-1}
\pdfoutput=1
\usepackage{amsmath}
\usepackage{braket}
\usepackage{graphicx}
\usepackage{hyperref}
\usepackage{color}

\begin{document}
\title{\color{blue}{DMRG-SCF study of the singlet, triplet, and quintet states of oxo-Mn(Salen)}}

\author{Sebastian Wouters}
\email{sebastianwouters@gmail.com}
\affiliation{Center for Molecular Modelling, Ghent University, Technologiepark 903, 9052 Zwijnaarde, Belgium}

\author{Thomas Bogaerts}
\affiliation{Center for Molecular Modelling, Ghent University, Technologiepark 903, 9052 Zwijnaarde, Belgium}
\affiliation{Center for Ordered Materials, Organometallics and Catalysis, Ghent University, Krijgslaan 281 (S3), 9000 Gent, Belgium}

\author{Pascal {Van Der Voort}}
\affiliation{Center for Ordered Materials, Organometallics and Catalysis, Ghent University, Krijgslaan 281 (S3), 9000 Gent, Belgium}

\author{Veronique {Van Speybroeck}}
\affiliation{Center for Molecular Modelling, Ghent University, Technologiepark 903, 9052 Zwijnaarde, Belgium}

\author{Dimitri {Van Neck}}
\affiliation{Center for Molecular Modelling, Ghent University, Technologiepark 903, 9052 Zwijnaarde, Belgium}

\begin{abstract}
We use \textsc{CheMPS2}, our free open-source spin-adapted implementation of the density matrix renormalization group (DMRG) [Wouters \textit{et al.}, \href{http://dx.doi.org/10.1016/j.cpc.2014.01.019}{\textit{Comput. Phys. Commun.} \textbf{185}, 1501 (2014)}], to study the lowest singlet, triplet, and quintet states of the oxo-Mn(Salen) complex. We describe how an initial approximate DMRG calculation in a large active space around the Fermi level can be used to obtain a good set of starting orbitals for subsequent complete-active-space or DMRG self-consistent field (CASSCF or DMRG-SCF) calculations. This procedure mitigates the need for a localization procedure, followed by a manual selection of the active space. Per multiplicity, the same active space of 28 electrons in 22 orbitals (28e, 22o) is obtained with the 6-31G*, cc-pVDZ, and ANO-RCC-VDZP basis sets (the latter with DKH2 scalar relativistic corrections). Our calculations provide new insight into the electronic structure of the quintet.
\end{abstract}

\maketitle

\begin{figure}[b!]
\includegraphics[width=0.2\textwidth]{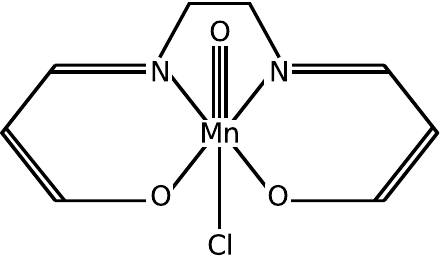}
\caption{\label{fig-structure} Model for the oxo-Mn(Salen) complex.}
\end{figure}

The manganese-salen complex is a high-yield catalyst for the enantioselective epoxidation of unfunctionalized olefins.\cite{Jacobsen, Irie19907345, JacobsenTwo, McGarrigle} Many density functional theory\cite{LindeDFT, StrassnerDFT, JacobsenDFT5,  *JacobsenDFT2, *JacobsenDFT3, *JacobsenDFT4, *JacobsenDFT6, *JacobsenDFT, *JacobsenDFT7, KhavrutskiiDFT1, *KhavrutskiiDFT2, *Khavrutskii20042004, *KhavrutskiiDFT5, *KhavrutskiiDFT4, AbashkinDFT2, EJOC200500042} and ab initio\cite{AbashkinDFT1, sears2, sears3} studies have tried to gain insight into its electronic structure and the energy barriers for possible reaction paths. A longstanding question in these studies is the relative stability of the singlet, triplet, and quintet states of the oxo-Mn(Salen) intermediate. This question has recently been addressed in several ab initio multireference (MR) studies,\cite{Ivanic2, Sears, Gagliardi} using the model in Fig. \ref{fig-structure}. The singlet and triplet were found to be nearly degenerate, and about 40 kcal/mol more stable than the quintet. The latter is well described by a single determinant, while the former two have outspoken MR character.

Only the singlet geometry can be optimized at the complete active space self-consistent field (CASSCF) level.\cite{Ivanic2} The triplet and quintet states are unstable with respect to dissociation into an oxygen atom (triplet) and the Mn(Salen) complex (quintet). This dissociation is spin-forbidden for the singlet. The CASSCF(10e, 10o)/6-31G* optimized singlet geometry from Ivanic \textit{et al.}\cite{Ivanic2} was therefore used here, as in the previous MR studies.

The relative stability of the singlet and triplet states is still under debate. Several MR studies have been performed in different active spaces, with various basis sets, and with or without the inclusion of dynamic correlation and relativistic effects.\cite{Ivanic2, Sears, Gagliardi} The importance of relativistic effects for the relative energies has been estimated both with an effective core potential\cite{Ivanic2} and with the perturbational Cowan-Griffin operator,\cite{Sears} and was found to be smaller than 0.2 kcal/mol. The effect of dynamic correlation has been assessed by applying MR perturbation theory on top of the CASSCF wavefunction.\cite{Ivanic2, Gagliardi} The corresponding variations in relative energy were about 5 kcal/mol, and can therefore not be neglected. The basis set choice shifted relative CASSCF energies by as much as 1.3 kcal/mol.\cite{Sears}

\begin{figure*}[t!]
\centering
\includegraphics[width=0.80\textwidth]{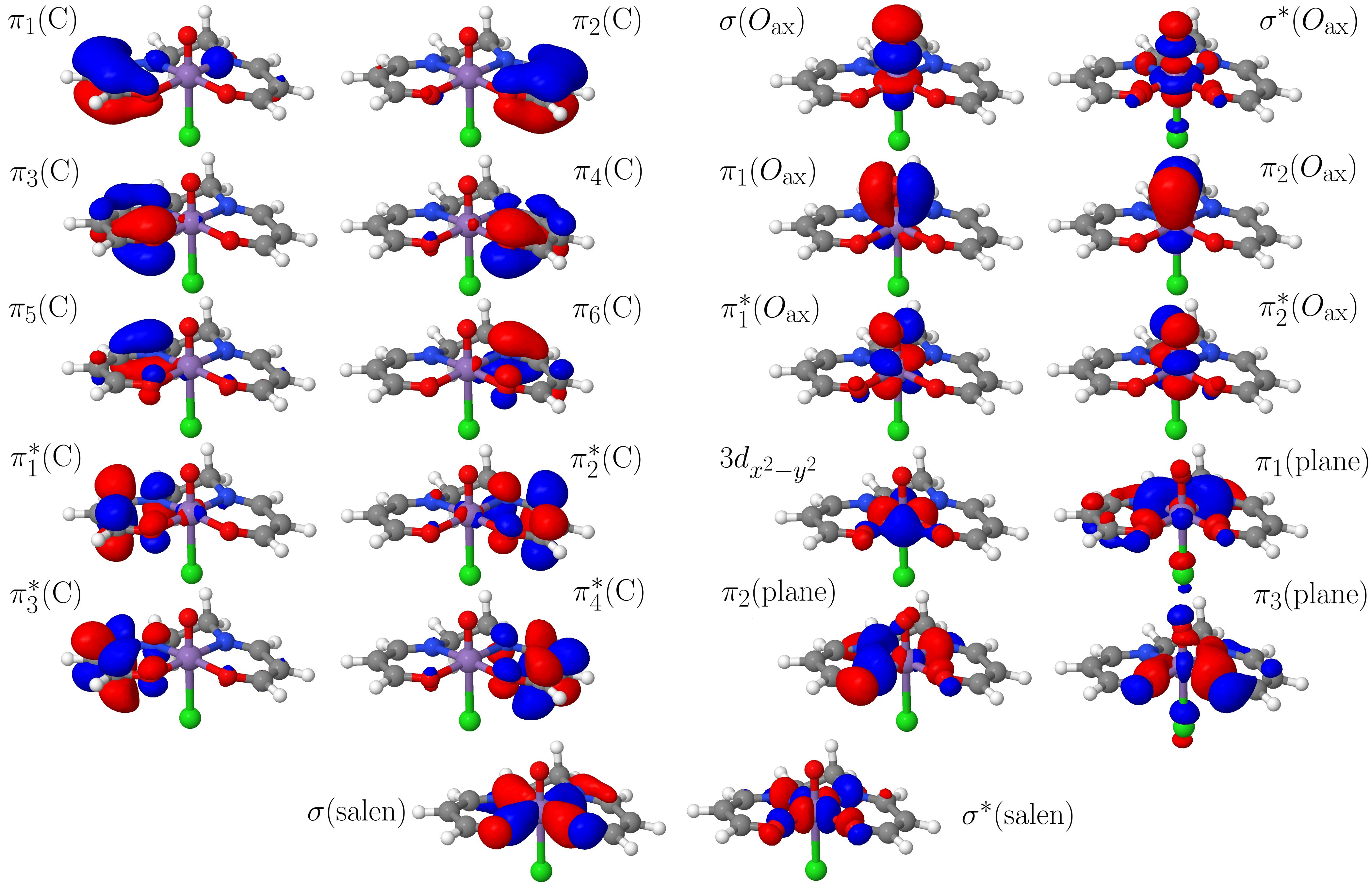}
\caption{\label{FIG-NOs} Natural orbitals of the converged (28e, 22o) singlet active space in the ANODZ basis. The same natural orbitals were found with the 6-31G* and cc-pVDZ basis sets.}
\end{figure*}

\begin{figure}[b!]
\centering
\includegraphics[width=0.32\textwidth]{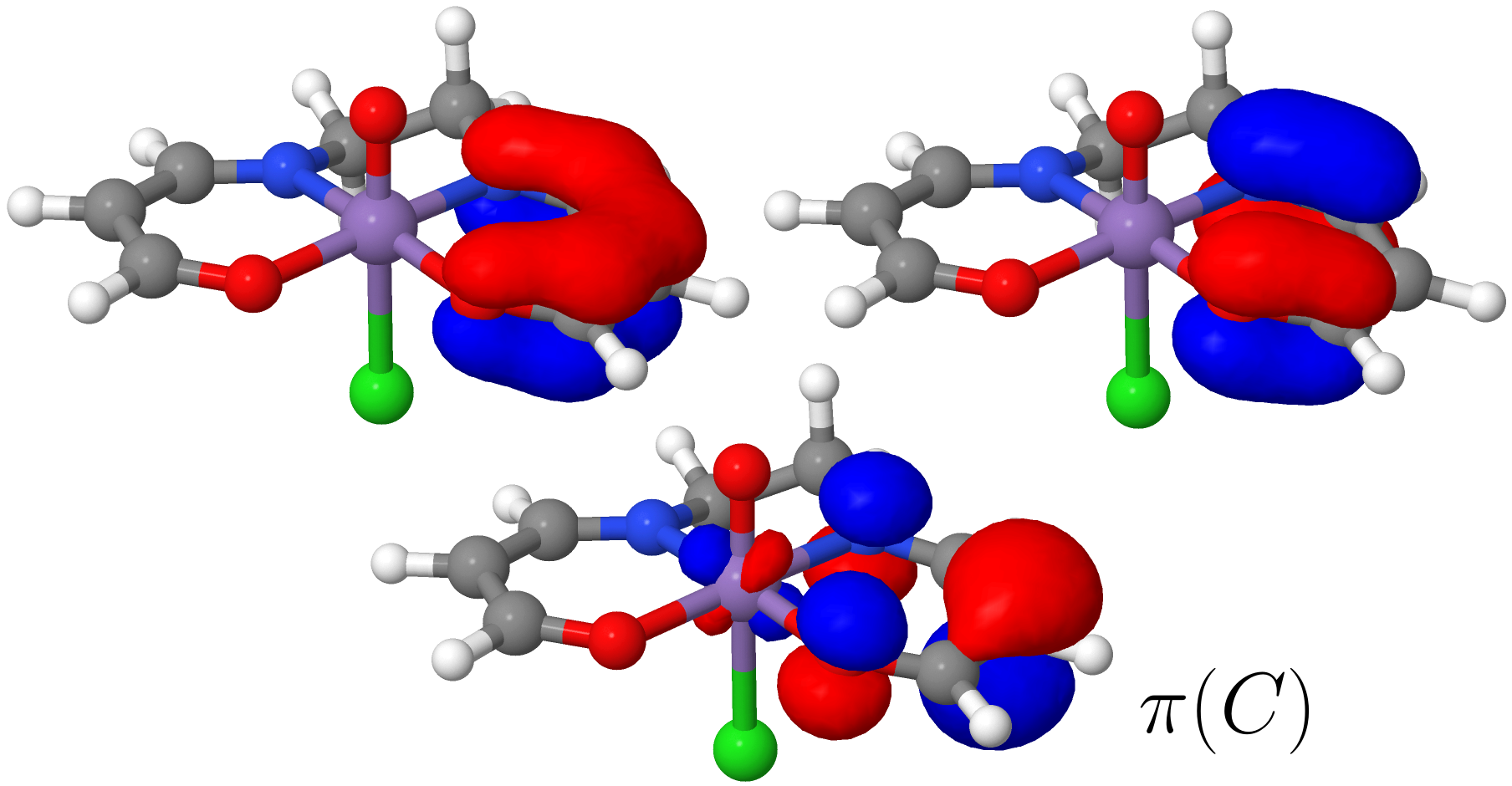}
\caption{\label{FIG-NOsQuintet} The three natural orbitals of the converged (28e, 22o) quintet active space in the ANODZ basis, which are not present in Fig. \ref{FIG-NOs}. NOON(top left) = 1.93; NOON(top right) = 1.89; and \textbf{NOON($\pi(C)$) = 1.01}. The same natural orbitals were found with the 6-31G* and cc-pVDZ basis sets.}
\end{figure}

In this work, we attempt to settle the debate on the active space selection for the geometry of Ref. \onlinecite{Ivanic2} with $\mathsf{C_1}$ symmetry (see Figs. \ref{FIG-NOs} and \ref{FIG-NOsQuintet}). Three double zeta basis sets with polarization functions are used in this work. The 6-31G* basis,\cite{631gstar1} also used in previous MR studies,\cite{Ivanic2, Sears} yields 273 (cartesian) orbitals. The cc-pVDZ basis\cite{ccpvdzbasis} has [$6s5p3d1f$] basis functions for Mn, and yields 293 (spherical) orbitals. And the ANO-RCC-VDZP basis\cite{anorccvdzp} with DKH2 scalar relativistic corrections\cite{DKH2citation} (\mbox{ANODZ}), also used in Ref. \onlinecite{Gagliardi}, yields 284 (spherical) orbitals. Restricted Hartree-Fock molecular orbitals were obtained with \textsc{Psi4}\cite{Psi4} for the 6-31G* and cc-pVDZ basis sets and with \textsc{Molpro}\cite{MOLPRO-WIREs} for \mbox{ANODZ}.

We now briefly describe the level of theory used in this work. For a more thorough discussion, we refer the reader to Refs. \onlinecite{ChanQCDMRG, CheMPS2paper}. The exact wavefunction in an active space of $L$ orbitals
\begin{eqnarray}
& \ket{\Psi} = \sum\limits_{\{n_{j\sigma}\}} C^{n_{1\uparrow} n_{1\downarrow} n_{2\uparrow} ... n_{L\downarrow}}  \nonumber \\
& \left( \hat{a}_{1\uparrow}^{\dagger} \right)^{ n_{1\uparrow} } \left( \hat{a}_{1\downarrow}^{\dagger} \right)^{ n_{1\downarrow} } \left( \hat{a}_{2\uparrow}^{\dagger} \right)^{ n_{2\uparrow} } ... \left( \hat{a}_{L\downarrow}^{\dagger} \right)^{ n_{L\downarrow} } \ket{-}, \label{FCIsolution}
\end{eqnarray}
grows exponentially fast (as $4^L$). One way to make computations tractable is by means of the density matrix renormalization group (DMRG).\cite{WhiteQCDMRG,  Mitrushenkov, ChanQCDMRG, PhysRevB.67.125114, ReiherCode, Zgid, KurashigeBasic, PhysRevB.81.235129, Sharma2012, wouters2012, CheMPS2paper} This method approximates the $C$-tensor of Eq. \eqref{FCIsolution} by a matrix product state (MPS):
\begin{eqnarray}
& C^{n_{1\uparrow} n_{1\downarrow} n_{2\uparrow} n_{2\downarrow} n_{3\uparrow} n_{3\downarrow} ... n_{L\uparrow} n_{L\downarrow}} = \nonumber \\
& \sum\limits_{\{ \alpha_k \}} A[1]^{n_{1\uparrow} n_{1\downarrow}}_{\alpha_1} ~ A[2]^{n_{2\uparrow} n_{2\downarrow}}_{\alpha_1;\alpha_2} ~ A[3]^{n_{3\uparrow} n_{3\downarrow}}_{\alpha_2;\alpha_3} ~ ... ~ A[L]^{n_{L\uparrow} n_{L\downarrow}}_{\alpha_{L-1}}. 
\end{eqnarray}
The indices $\alpha_k$ are called the \textit{bond} or \textit{virtual} indices. They have to grow exponentially towards the middle of the MPS chain to represent a general $C$-tensor. The exponential complexity is removed when their rank is truncated to a fixed virtual dimension $D$: $\text{dim}(\alpha_k) = \min(4^k, 4^{L-k}, D)$.
By properly exploiting the gauge freedom of the MPS ansatz,\cite{CheMPS2paper} the simultaneous optimization of two neighbouring MPS site tensors can always be written as a numerically stable standard Hermitian eigenvalue problem. The DMRG algorithm sweeps back and forth through the chain, while locally optimizing the MPS site tensors, until energy and/or wavefunction convergence is reached.\cite{ChanQCDMRG,CheMPS2paper}

\begin{figure}[t!]
\centering
\includegraphics[width=0.44\textwidth]{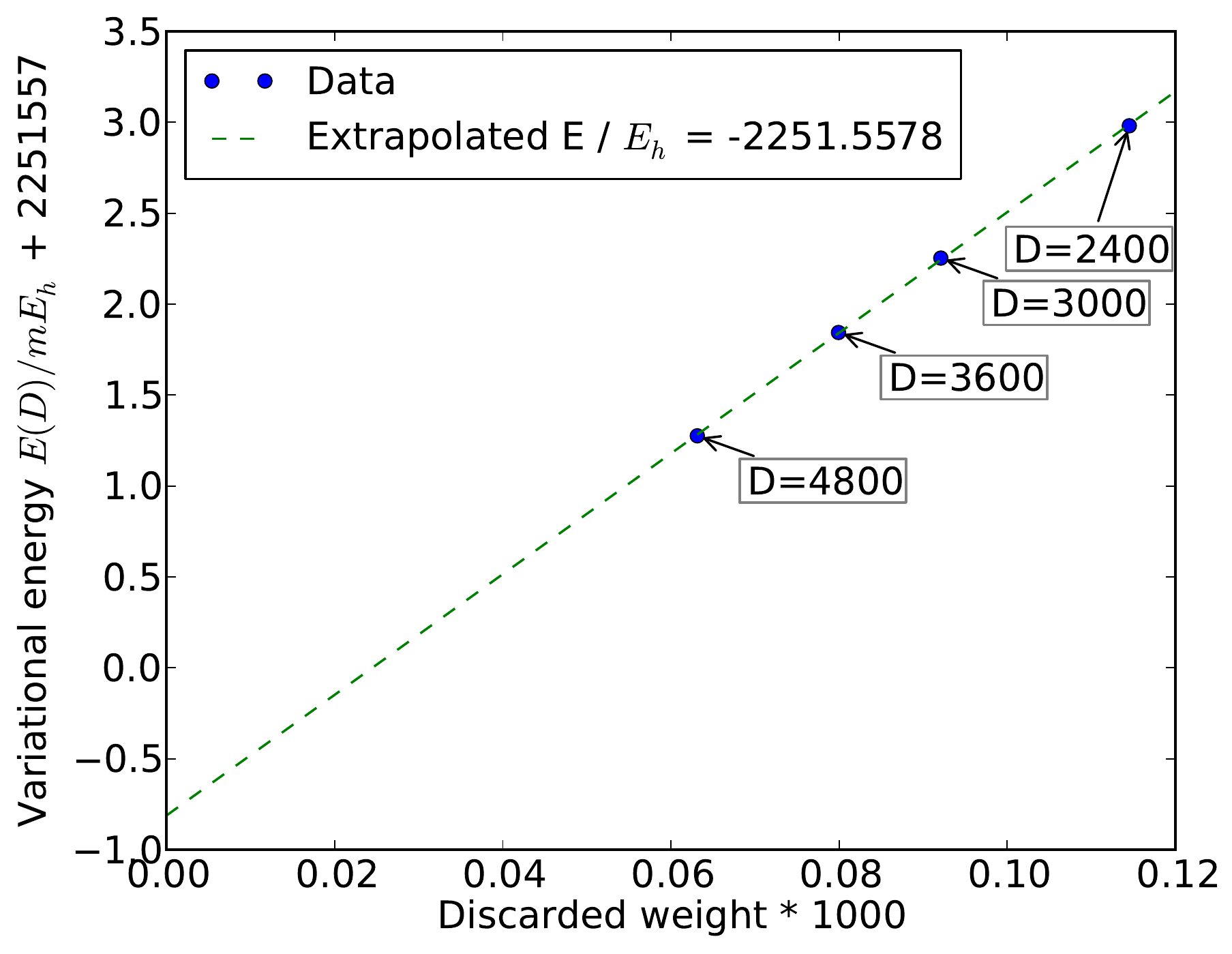}
\caption{\label{FIG-Extrapol} Energy extrapolation for the converged (28e, 22o) active space of the triplet in the 6-31G* basis. $D$ denotes the number of \textit{reduced} virtual basis states.}
\end{figure}

The so-called discarded weight is a nonnegative measure which indicates the aptitude of an MPS to represent the exact solution.\cite{ChanQCDMRG,CheMPS2paper}
Both the variational energy and the discarded weight decrease with increasing virtual dimension. A linear extrapolation between both allows to estimate the exact ground state energy,\cite{ChanQCDMRG, CheMPS2paper, PhysRevB.53.14349} see Fig. \ref{FIG-Extrapol}.

To reduce the computational cost, as well as to be able to tackle different symmetry sectors separately, symmetry-adapted MPS are often used.\cite{Sharma2012,CheMPS2paper} \textsc{CheMPS2}, our free open-source spin-adapted implementation of DMRG,\cite{CheMPS2github, CheMPS2paper} exploits $\mathsf{SU(2)}$ spin symmetry, $\mathsf{U(1)}$ particle-number symmetry, and the abelian point groups with real-valued character tables.\footnote{Each spin multiplet in a spin-adapted MPS is represented by only one reduced basis state. This is the \textit{reduced} to which is referred in the caption of Fig. \ref{FIG-Extrapol}.} We can therefore calculate the lowest singlet, triplet, and quintet states of the oxo-Mn(Salen) complex as three ground-state calculations in different symmetry sectors.

In methods which use a full-configuration-interaction (FCI) solver, this solver can be replaced by DMRG. DMRG allows for an efficient extraction of the reduced two-body density matrix (2-RDM).\cite{Zgid-2RDM, Ghosh-DMRGSCF} The 2-RDM of the active space is required in the CASSCF method to compute the gradient and the Hessian with respect to orbital rotations.\cite{RDMbasedCASSCF} It is therefore natural to introduce a CASSCF variant with DMRG as active space solver, called DMRG-SCF,\cite{Zgid-DMRGSCF, Ghosh-DMRGSCF, QUA22099} which allows to treat static correlation in large active spaces. In \textsc{CheMPS2}, we have implemented the augmented Hessian Newton-Raphson DMRG-SCF method, with exact Hessian.\cite{RDMbasedCASSCF,CheMPS2paper}

DMRG is an ideal candidate to study the electronic structure of transition metal systems, as they typically have large active spaces. Reiher and coworkers realized this capability of DMRG, and identified $\text{Cr}_2$ and $[\text{Cu}_2\text{O}_2]^{2+}$ as interesting cases.\cite{ReiherCode, Reiher2, Reiher3} Yanai and coworkers were eventually able to fully resolve their potential energy surfaces.\cite{KurashigeBasic, DMRG-CT, DMRG-CASPT2} This has triggered many interesting DMRG studies of transition metal systems.\cite{Sharma2012, ct300211j, jz301319v, naturechem, 1.4863345, C3CP55225J}

For CASSCF calculations, an initial active space is required. It is often constructed by localizing the occupied and virtual molecular orbitals separately, and by manually selecting an interesting subset.\cite{Ivanic2, Sears} However, this subset can be biased, and might converge to a local minimum. Conversely, orbitals with occupation numbers far from empty of filled, lie close to the Fermi level.\cite{Bible} To bypass the localization procedure and the manual selection of the active space, we have performed approximate DMRG calculations for the singlet in a large window around the Fermi level (with $D_{\mathsf{SU(2)}}=2000$ the \textit{reduced} virtual dimension). The window was chosen based on the shapes of the molecular orbitals: it should include at least (in rotated form) the active space of Ref. \onlinecite{Gagliardi}. For the basis sets 6-31G*, cc-pVDZ, and ANODZ, the active space window had the size (50e, 40o); (50e, 44o); and (56e, 45o), respectively. From the approximate DMRG calculation, the natural orbitals with occupation number (NOON) in the range 0.015 to 1.985 were kept, and used for the subsequent DMRG-SCF singlet calculations (with $D_{\mathsf{SU(2)}}=3000$). Remarkably, with the three basis sets the same (28e, 22o) active space was retrieved. When the 2-norm of the gradient was smaller than $10^{-4}$, the DMRG-SCF calculations were branched to calculate the triplet and quintet as well. After convergence of the active spaces, a larger DMRG calculation with $D_{\mathsf{SU(2)}}=4800$ was performed to extrapolate the variational energies to the exact result, see Fig. \ref{FIG-Extrapol}. For the DMRG calculations, the natural orbitals were used, and they were ordered according to the NOON.\footnote{Using natural orbitals, and ordering them according to the NOON, is not an optimal choice for DMRG. It is better to group corresponding bonding and antibonding orbitals. However, this procedure allows for an unmonitored optimization.}

The converged (28e, 22o) singlet active space in the ANODZ basis is shown in Fig. \ref{FIG-NOs}. The same (28e, 22o) singlet active space is found with the 6-31G* and cc-pVDZ basis sets. The (18e, 17o) active space of Ref. \onlinecite{Gagliardi} is augmented in Fig. \ref{FIG-NOs} with two extra $\pi$-orbitals for the conjugated system, so that the active space now contains a $\pi$-orbital per participating atom. In addition, the nonbonding $3d_{x^2-y^2}$ orbital of Mn interacts with the in-plane $\pi$-orbitals of the oxygen and nitrogen atoms of the Salen ligand, further augmenting the active space with three extra orbitals. The NOON are listed in Tab. \ref{Tab-NOON}. The triplet has the same natural orbitals as the singlet, and can roughly be interpreted as the electron excitation $3d_{x^2-y^2} \rightarrow \pi_1^*(O_{\text{ax}})$ from the singlet, as noted by Ref. \onlinecite{Ivanic2}. In the quintet, the $\pi_2(C)$, $\pi_4(C)$, and $\pi_6(C)$ orbitals are rotated into new natural orbitals, which are shown in Fig. \ref{FIG-NOsQuintet} together with their NOON. The same rotated quintet natural orbitals are found with the 6-31G* and cc-pVDZ basis sets. The quintet can hence be interpreted as the additional electron excitation $\pi(C) \rightarrow \pi_2^*(O_{\text{ax}})$ from the triplet. The electronic structure of the quintet differs from previous studies,\cite{Ivanic2, Sears} where it was identified as the electron excitation $\pi_2(O_{\text{ax}}) \rightarrow \pi_2^*(O_{\text{ax}})$ from the triplet.

\begin{table}[b!]
\caption{ \label{Tab-NOON} NOON of the converged (28e, 22o) singlet ($^1A$), triplet ($^3A$), and quintet ($^5A$) active spaces in the ANODZ basis. The natural orbitals $\pi_2(C)$, $\pi_4(C)$, and $\pi_6(C)$ are rotated for the quintet; they are given in Fig. \ref{FIG-NOsQuintet}.}
\begin{tabular}{l rrr l rrr}
\hline
\hline
 & $^1A$ & $^3A$ & $^5A$ ~ &  & ~ $^1A$ & $^3A$ & $^5A$\\
\hline
$\pi_1(C)$                & 1.99 & 1.99 & 1.99 ~ & ~ $\sigma(O_{\text{ax}})$   & 1.91 & 1.90 & 1.89 \\
$\pi_2(C)$                & 1.99 & 1.99 & -    ~ & ~ $\sigma^*(O_{\text{ax}})$ & 0.11 & 0.11 & 0.12 \\
$\pi_3(C)$                & 1.96 & 1.96 & 1.96 ~ & ~ $\pi_1(O_{\text{ax}})$    & 1.86 & 1.77 & 1.94 \\
$\pi_4(C)$                & 1.96 & 1.96 & -    ~ & ~ $\pi_2(O_{\text{ax}})$    & 1.85 & 1.95 & 1.94 \\
$\pi_5(C)$                & 1.94 & 1.94 & 1.94 ~ & ~ $\pi_1^*(O_{\text{ax}})$  & \textbf{0.17} & \textbf{1.04} & \textbf{1.05} \\
$\pi_6(C)$                & 1.94 & 1.94 & -    ~ & ~ $\pi_2^*(O_{\text{ax}})$  & \textbf{0.17} & \textbf{0.24} & \textbf{1.04} \\
$\pi_1^*(C)$              & 0.07 & 0.07 & 0.07 ~ & ~ $3d_{x^2-y^2}$            & \textbf{1.97} & \textbf{1.00} & \textbf{1.00} \\
$\pi_2^*(C)$              & 0.07 & 0.07 & 0.11 ~ & ~ $\pi_1(\text{plane})$     & 1.99 & 1.99 & 1.99 \\
$\pi_3^*(C)$              & 0.03 & 0.03 & 0.03 ~ & ~ $\pi_2(\text{plane})$     & 1.98 & 1.98 & 1.99 \\
$\pi_4^*(C)$              & 0.03 & 0.03 & 0.06 ~ & ~ $\pi_3(\text{plane})$     & 1.98 & 1.98 & 1.99 \\
$\sigma(\text{salen})$    & 1.95 & 1.93 & 1.98 ~ & ~ $\sigma^*(\text{salen})$  & 0.08 & 0.10 & 0.07 \\
\hline
\hline
\end{tabular}
\end{table}

Our study found a different electronic structure for the quintet due to the larger active space for the conjugated system. Refs. \onlinecite{Ivanic2, Sears} only considered two bonding and two antibonding orbitals for the conjugated system, i.e. four $\pi$-orbitals instead of the ten $\pi$-orbitals in our study. (Ref. \onlinecite{Gagliardi} did not study the quintet.) In the quintet, the NOON of all five $\pi$-orbitals on the right wing of the structure deviate significantly compared to the singlet and triplet, see Tab. \ref{Tab-NOON} and Fig. \ref{FIG-NOsQuintet}. It is hence crucial for the description of the electronic structure of the quintet to incorporate these five $\pi$-orbitals in the active space. To have an equilibrated description, the corresponding orbitals on the left wing should also be included, requiring a total of ten $\pi$-orbitals. In hindsight, the inclusion of the three in-plane $\pi_{\{1,2,3\}}(\text{plane})$ orbitals was not necessary, given the NOON in Tab. \ref{Tab-NOON}.

In the simplified single determinant picture based on Tab. \ref{Tab-NOON}, in which the quintet ground state corresponds to 
\begin{equation}
\ket{\left[ \pi(C) \right]^{\uparrow} \left[\pi_2(O_{\text{ax}})\right]^{\uparrow\downarrow} \left[3d_{x^2-y^2}\right]^{\uparrow} \left[\pi_1^*(O_{\text{ax}})\right]^{\uparrow} \left[\pi_2^*(O_{\text{ax}})\right]^{\uparrow} }, \label{QuintetGSdeterminant}
\end{equation}
we can interpret the quintet from Refs. \onlinecite{Ivanic2, Sears}
\begin{equation}
\ket{\left[ \pi(C) \right]^{\uparrow\downarrow} \left[\pi_2(O_{\text{ax}})\right]^{\uparrow} \left[3d_{x^2-y^2}\right]^{\uparrow} \left[\pi_1^*(O_{\text{ax}})\right]^{\uparrow} \left[\pi_2^*(O_{\text{ax}})\right]^{\uparrow} }
\end{equation}
as the single electron excitation $\pi_2(O_{\text{ax}}) \rightarrow \pi(C)$ from the quintet ground state determinant \eqref{QuintetGSdeterminant}. The reason why Refs. \onlinecite{Ivanic2, Sears} found a different quintet ground state determinant can then be recast to: (a part of) the $\pi(C)$ orbital was explicitly kept doubly occupied.

\begin{table}[t!]
\caption{ \label{Tab-Absolute} Absolute energies in Hartree and relative energies in kcal/mol; obtained by extrapolating the DMRG-SCF(28e, 22o) energies with discarded weight. In square brackets the CASSCF(12e, 11o)/6-31G* and GASSCF(18e, 17o)/ANODZ results from Refs. \onlinecite{Sears} and \onlinecite{Gagliardi} are given for comparison.}
\begin{tabular}{llll}
\hline
\hline
& ~ 6-31G* & ~ cc-pVDZ & ~ ANODZ\\
\hline
E($^1A$)            & ~ -2251.5498 & ~ -2251.7509 & ~ -2261.0226\\
E($^3A$)            & ~ -2251.5578 & ~ -2251.7593 & ~ -2261.0290\\
E($^5A$)            & ~ -2251.5268 & ~ -2251.7316 & ~ -2260.9994\\
E($^3A$) - E($^1A$) & ~ -5.0  [0.3]& ~ -5.3       & ~ -4.0 [-3.6]\\
E($^5A$) - E($^1A$) & ~ 14.5 [42.9]& ~ 12.1       & ~ 14.5\\
\hline
\hline

\end{tabular}
\end{table}

The energies for the different multiplicities and basis sets are given in Tab. \ref{Tab-Absolute}. The energies are consistent for the three basis sets studied. The triplet has the lowest energy. The quintet lies only 12-14 kcal/mol above the singlet, much lower than what was observed in a smaller active space.\cite{Ivanic2, Sears} Note that the addition of dynamic correlation can still shift the relative energies by $\sim$5 kcal/mol.\cite{Ivanic2,Gagliardi}

In conclusion, we have studied the active spaces and the relative stability of the lowest singlet, triplet, and quintet states of the oxo-Mn(salen) complex. With an initial approximate DMRG calculation in a large window around the Fermi level, we have obtained a good set of starting orbitals for the DMRG-SCF calculations, \textit{without} an explicit localization procedure and subsequent manual selection of the active space. Per multiplicity, the same active space was obtained with the basis sets 6-31G*, cc-pVDZ, and ANODZ. The electronic structure of the quintet differs from previous studies. It can be interpreted by the two-electron excitation $3d_{x^2-y^2} \rightarrow \pi_1^*(O_{\text{ax}})$ and $\pi(C) \rightarrow \pi_2^*(O_{\text{ax}})$ from the singlet. We found that the triplet is 5 kcal/mol more stable than the singlet, and that the quintet lies only 12-14 kcal/mol higher than the singlet. In the future, we would like to study the experimental structure,\cite{Jacobsen} use a triple zeta basis, and add dynamic correlation. The experimental structure has a larger $\pi$-conjugated system, in which the quintet electron excitation takes place. Dynamic correlation can be added with perturbation theory (DMRG-CASPT2),\cite{C3CP55225J, DMRG-CASPT2} configuration interaction (DMRG-MRCI),\cite{C3CP55225J, DMRG-MRCI} or canonical transformations (DMRG-CT).\cite{DMRG-CT} We would also like to study reactions with the Mn(Salen) catalyst.\cite{C3CC44473B}

S.W. acknowledges a Ph.D. fellowship from the Research Foundation Flanders. The research was financed by UGent GOA grant 01G00710, and by the European Research Council through the European Community's Seventh Framework Programme FP7(2007-2013), ERC grant 240483. The computational resources (Stevin Supercomputer Infrastructure) and services used in this work were provided by the VSC (Flemish Supercomputer Center), funded by Ghent University, the Hercules Foundation and the Flemish Government - department EWI.

\bibliography{biblio}

\begin{thebibliography}{64}%
\makeatletter
\providecommand \@ifxundefined [1]{%
 \@ifx{#1\undefined}
}%
\providecommand \@ifnum [1]{%
 \ifnum #1\expandafter \@firstoftwo
 \else \expandafter \@secondoftwo
 \fi
}%
\providecommand \@ifx [1]{%
 \ifx #1\expandafter \@firstoftwo
 \else \expandafter \@secondoftwo
 \fi
}%
\providecommand \natexlab [1]{#1}%
\providecommand \enquote  [1]{``#1''}%
\providecommand \bibnamefont  [1]{#1}%
\providecommand \bibfnamefont [1]{#1}%
\providecommand \citenamefont [1]{#1}%
\providecommand \href@noop [0]{\@secondoftwo}%
\providecommand \href [0]{\begingroup \@sanitize@url \@href}%
\providecommand \@href[1]{\@@startlink{#1}\@@href}%
\providecommand \@@href[1]{\endgroup#1\@@endlink}%
\providecommand \@sanitize@url [0]{\catcode `\\12\catcode `\$12\catcode
  `\&12\catcode `\#12\catcode `\^12\catcode `\_12\catcode `\%12\relax}%
\providecommand \@@startlink[1]{}%
\providecommand \@@endlink[0]{}%
\providecommand \url  [0]{\begingroup\@sanitize@url \@url }%
\providecommand \@url [1]{\endgroup\@href {#1}{\urlprefix }}%
\providecommand \urlprefix  [0]{URL }%
\providecommand \Eprint [0]{\href }%
\providecommand \doibase [0]{http://dx.doi.org/}%
\providecommand \selectlanguage [0]{\@gobble}%
\providecommand \bibinfo  [0]{\@secondoftwo}%
\providecommand \bibfield  [0]{\@secondoftwo}%
\providecommand \translation [1]{[#1]}%
\providecommand \BibitemOpen [0]{}%
\providecommand \bibitemStop [0]{}%
\providecommand \bibitemNoStop [0]{.\EOS\space}%
\providecommand \EOS [0]{\spacefactor3000\relax}%
\providecommand \BibitemShut  [1]{\csname bibitem#1\endcsname}%
\let\auto@bib@innerbib\@empty
\bibitem [{\citenamefont {Zhang}\ \emph {et~al.}(1990)\citenamefont {Zhang},
  \citenamefont {Loebach}, \citenamefont {Wilson},\ and\ \citenamefont
  {Jacobsen}}]{Jacobsen}%
  \BibitemOpen
  \bibfield  {author} {\bibinfo {author} {\bibfnamefont {W.}~\bibnamefont
  {Zhang}}, \bibinfo {author} {\bibfnamefont {J.~L.}\ \bibnamefont {Loebach}},
  \bibinfo {author} {\bibfnamefont {S.~R.}\ \bibnamefont {Wilson}}, \ and\
  \bibinfo {author} {\bibfnamefont {E.~N.}\ \bibnamefont {Jacobsen}},\ }\href
  {\doibase 10.1021/ja00163a052} {\bibfield  {journal} {\bibinfo  {journal} {J.
  Am. Chem. Soc.}\ }\textbf {\bibinfo {volume} {112}},\ \bibinfo {pages} {2801}
  (\bibinfo {year} {1990})}\BibitemShut {NoStop}%
\bibitem [{\citenamefont {Irie}\ \emph {et~al.}(1990)\citenamefont {Irie},
  \citenamefont {Noda}, \citenamefont {Ito}, \citenamefont {Matsumoto},\ and\
  \citenamefont {Katsuki}}]{Irie19907345}%
  \BibitemOpen
  \bibfield  {author} {\bibinfo {author} {\bibfnamefont {R.}~\bibnamefont
  {Irie}}, \bibinfo {author} {\bibfnamefont {K.}~\bibnamefont {Noda}}, \bibinfo
  {author} {\bibfnamefont {Y.}~\bibnamefont {Ito}}, \bibinfo {author}
  {\bibfnamefont {N.}~\bibnamefont {Matsumoto}}, \ and\ \bibinfo {author}
  {\bibfnamefont {T.}~\bibnamefont {Katsuki}},\ }\href {\doibase
  10.1016/S0040-4039(00)88562-7} {\bibfield  {journal} {\bibinfo  {journal}
  {Tetrahedron Lett.}\ }\textbf {\bibinfo {volume} {31}},\ \bibinfo {pages}
  {7345 } (\bibinfo {year} {1990})}\BibitemShut {NoStop}%
\bibitem [{\citenamefont {Jacobsen}\ \emph {et~al.}(1991)\citenamefont
  {Jacobsen}, \citenamefont {Zhang}, \citenamefont {Muci}, \citenamefont
  {Ecker},\ and\ \citenamefont {Deng}}]{JacobsenTwo}%
  \BibitemOpen
  \bibfield  {author} {\bibinfo {author} {\bibfnamefont {E.~N.}\ \bibnamefont
  {Jacobsen}}, \bibinfo {author} {\bibfnamefont {W.}~\bibnamefont {Zhang}},
  \bibinfo {author} {\bibfnamefont {A.~R.}\ \bibnamefont {Muci}}, \bibinfo
  {author} {\bibfnamefont {J.~R.}\ \bibnamefont {Ecker}}, \ and\ \bibinfo
  {author} {\bibfnamefont {L.}~\bibnamefont {Deng}},\ }\href {\doibase
  10.1021/ja00018a068} {\bibfield  {journal} {\bibinfo  {journal} {J. Am. Chem.
  Soc.}\ }\textbf {\bibinfo {volume} {113}},\ \bibinfo {pages} {7063} (\bibinfo
  {year} {1991})}\BibitemShut {NoStop}%
\bibitem [{\citenamefont {McGarrigle}\ and\ \citenamefont
  {Gilheany}(2005)}]{McGarrigle}%
  \BibitemOpen
  \bibfield  {author} {\bibinfo {author} {\bibfnamefont {E.~M.}\ \bibnamefont
  {McGarrigle}}\ and\ \bibinfo {author} {\bibfnamefont {D.~G.}\ \bibnamefont
  {Gilheany}},\ }\href {\doibase 10.1021/cr0306945} {\bibfield  {journal}
  {\bibinfo  {journal} {Chem. Rev.}\ }\textbf {\bibinfo {volume} {105}},\
  \bibinfo {pages} {1563} (\bibinfo {year} {2005})}\BibitemShut {NoStop}%
\bibitem [{\citenamefont {Linde}\ \emph {et~al.}(1999)\citenamefont {Linde},
  \citenamefont {\r{A}kermark}, \citenamefont {Norrby},\ and\ \citenamefont
  {Svensson}}]{LindeDFT}%
  \BibitemOpen
  \bibfield  {author} {\bibinfo {author} {\bibfnamefont {C.}~\bibnamefont
  {Linde}}, \bibinfo {author} {\bibfnamefont {B.}~\bibnamefont {\r{A}kermark}},
  \bibinfo {author} {\bibfnamefont {P.-O.}\ \bibnamefont {Norrby}}, \ and\
  \bibinfo {author} {\bibfnamefont {M.}~\bibnamefont {Svensson}},\ }\href
  {\doibase 10.1021/ja9809915} {\bibfield  {journal} {\bibinfo  {journal} {J.
  Am. Chem. Soc.}\ }\textbf {\bibinfo {volume} {121}},\ \bibinfo {pages} {5083}
  (\bibinfo {year} {1999})}\BibitemShut {NoStop}%
\bibitem [{\citenamefont {Strassner}\ and\ \citenamefont
  {Houk}(1999)}]{StrassnerDFT}%
  \BibitemOpen
  \bibfield  {author} {\bibinfo {author} {\bibfnamefont {T.}~\bibnamefont
  {Strassner}}\ and\ \bibinfo {author} {\bibfnamefont {K.~N.}\ \bibnamefont
  {Houk}},\ }\href {\doibase 10.1021/ol990064i} {\bibfield  {journal} {\bibinfo
   {journal} {Org. Lett.}\ }\textbf {\bibinfo {volume} {1}},\ \bibinfo {pages}
  {419} (\bibinfo {year} {1999})}\BibitemShut {NoStop}%
\bibitem [{\citenamefont {Cavallo}\ and\ \citenamefont
  {Jacobsen}(2000)}]{JacobsenDFT5}%
  \BibitemOpen
  \bibfield  {author} {\bibinfo {author} {\bibfnamefont {L.}~\bibnamefont
  {Cavallo}}\ and\ \bibinfo {author} {\bibfnamefont {H.}~\bibnamefont
  {Jacobsen}},\ }\href {\doibase
  10.1002/(SICI)1521-3773(20000204)39:3<589::AID-ANIE589>3.0.CO;2-0} {\bibfield
   {journal} {\bibinfo  {journal} {Angew. Chem. Int. Ed. (English)}\ }\textbf
  {\bibinfo {volume} {39}},\ \bibinfo {pages} {589} (\bibinfo {year}
  {2000})}\BibitemShut {NoStop}%
\bibitem [{\citenamefont {Cavallo}\ and\ \citenamefont
  {Jacobsen}(2003{\natexlab{a}})}]{JacobsenDFT2}%
  \BibitemOpen
  \bibfield  {author} {\bibinfo {author} {\bibfnamefont {L.}~\bibnamefont
  {Cavallo}}\ and\ \bibinfo {author} {\bibfnamefont {H.}~\bibnamefont
  {Jacobsen}},\ }\href {\doibase 10.1021/jo034059a} {\bibfield  {journal}
  {\bibinfo  {journal} {J. Org. Chem.}\ }\textbf {\bibinfo {volume} {68}},\
  \bibinfo {pages} {6202} (\bibinfo {year} {2003}{\natexlab{a}})}\BibitemShut
  {NoStop}%
\bibitem [{\citenamefont {Cavallo}\ and\ \citenamefont
  {Jacobsen}(2003{\natexlab{b}})}]{JacobsenDFT3}%
  \BibitemOpen
  \bibfield  {author} {\bibinfo {author} {\bibfnamefont {L.}~\bibnamefont
  {Cavallo}}\ and\ \bibinfo {author} {\bibfnamefont {H.}~\bibnamefont
  {Jacobsen}},\ }\href {\doibase 10.1021/jp034194r} {\bibfield  {journal}
  {\bibinfo  {journal} {J. Phys. Chem. A}\ }\textbf {\bibinfo {volume} {107}},\
  \bibinfo {pages} {5466} (\bibinfo {year} {2003}{\natexlab{b}})}\BibitemShut
  {NoStop}%
\bibitem [{\citenamefont {Cavallo}\ and\ \citenamefont
  {Jacobsen}(2003{\natexlab{c}})}]{JacobsenDFT4}%
  \BibitemOpen
  \bibfield  {author} {\bibinfo {author} {\bibfnamefont {L.}~\bibnamefont
  {Cavallo}}\ and\ \bibinfo {author} {\bibfnamefont {H.}~\bibnamefont
  {Jacobsen}},\ }\href {\doibase 10.1002/ejic.200390118} {\bibfield  {journal}
  {\bibinfo  {journal} {Eur. J. Inorg. Chem.}\ }\textbf {\bibinfo {volume}
  {2003}},\ \bibinfo {pages} {892} (\bibinfo {year}
  {2003}{\natexlab{c}})}\BibitemShut {NoStop}%
\bibitem [{\citenamefont {Cavallo}\ and\ \citenamefont
  {Jacobsen}(2004)}]{JacobsenDFT6}%
  \BibitemOpen
  \bibfield  {author} {\bibinfo {author} {\bibfnamefont {L.}~\bibnamefont
  {Cavallo}}\ and\ \bibinfo {author} {\bibfnamefont {H.}~\bibnamefont
  {Jacobsen}},\ }\href {\doibase 10.1021/ic0353615} {\bibfield  {journal}
  {\bibinfo  {journal} {Inorg. Chem.}\ }\textbf {\bibinfo {volume} {43}},\
  \bibinfo {pages} {2175} (\bibinfo {year} {2004})}\BibitemShut {NoStop}%
\bibitem [{\citenamefont {Jacobsen}\ and\ \citenamefont
  {Cavallo}(2001)}]{JacobsenDFT}%
  \BibitemOpen
  \bibfield  {author} {\bibinfo {author} {\bibfnamefont {H.}~\bibnamefont
  {Jacobsen}}\ and\ \bibinfo {author} {\bibfnamefont {L.}~\bibnamefont
  {Cavallo}},\ }\href {\doibase
  10.1002/1521-3765(20010216)7:4<800::AID-CHEM800>3.0.CO;2-1} {\bibfield
  {journal} {\bibinfo  {journal} {Chem. Eur. J.}\ }\textbf {\bibinfo {volume}
  {7}},\ \bibinfo {pages} {800} (\bibinfo {year} {2001})}\BibitemShut {NoStop}%
\bibitem [{\citenamefont {Jacobsen}\ and\ \citenamefont
  {Cavallo}(2004)}]{JacobsenDFT7}%
  \BibitemOpen
  \bibfield  {author} {\bibinfo {author} {\bibfnamefont {H.}~\bibnamefont
  {Jacobsen}}\ and\ \bibinfo {author} {\bibfnamefont {L.}~\bibnamefont
  {Cavallo}},\ }\href {\doibase 10.1039/B402188F} {\bibfield  {journal}
  {\bibinfo  {journal} {Phys. Chem. Chem. Phys.}\ }\textbf {\bibinfo {volume}
  {6}},\ \bibinfo {pages} {3747} (\bibinfo {year} {2004})}\BibitemShut
  {NoStop}%
\bibitem [{\citenamefont {Khavrutskii}, \citenamefont {Musaev},\ and\
  \citenamefont {Morokuma}(2003{\natexlab{a}})}]{KhavrutskiiDFT1}%
  \BibitemOpen
  \bibfield  {author} {\bibinfo {author} {\bibfnamefont {I.~V.}\ \bibnamefont
  {Khavrutskii}}, \bibinfo {author} {\bibfnamefont {D.~G.}\ \bibnamefont
  {Musaev}}, \ and\ \bibinfo {author} {\bibfnamefont {K.}~\bibnamefont
  {Morokuma}},\ }\href {\doibase 10.1021/ic026094q} {\bibfield  {journal}
  {\bibinfo  {journal} {Inorg. Chem.}\ }\textbf {\bibinfo {volume} {42}},\
  \bibinfo {pages} {2606} (\bibinfo {year} {2003}{\natexlab{a}})}\BibitemShut
  {NoStop}%
\bibitem [{\citenamefont {Khavrutskii}, \citenamefont {Musaev},\ and\
  \citenamefont {Morokuma}(2003{\natexlab{b}})}]{KhavrutskiiDFT2}%
  \BibitemOpen
  \bibfield  {author} {\bibinfo {author} {\bibfnamefont {I.~V.}\ \bibnamefont
  {Khavrutskii}}, \bibinfo {author} {\bibfnamefont {D.~G.}\ \bibnamefont
  {Musaev}}, \ and\ \bibinfo {author} {\bibfnamefont {K.}~\bibnamefont
  {Morokuma}},\ }\href {\doibase 10.1021/ja0343656} {\bibfield  {journal}
  {\bibinfo  {journal} {J. Am. Chem. Soc.}\ }\textbf {\bibinfo {volume}
  {125}},\ \bibinfo {pages} {13879} (\bibinfo {year}
  {2003}{\natexlab{b}})}\BibitemShut {NoStop}%
\bibitem [{\citenamefont {Khavrutskii}, \citenamefont {Musaev},\ and\
  \citenamefont {Morokuma}(2004)}]{Khavrutskii20042004}%
  \BibitemOpen
  \bibfield  {author} {\bibinfo {author} {\bibfnamefont {I.~V.}\ \bibnamefont
  {Khavrutskii}}, \bibinfo {author} {\bibfnamefont {D.~G.}\ \bibnamefont
  {Musaev}}, \ and\ \bibinfo {author} {\bibfnamefont {K.}~\bibnamefont
  {Morokuma}},\ }\href {\doibase 10.1073/pnas.0307082101} {\bibfield  {journal}
  {\bibinfo  {journal} {Proc. Natl. Acad. Sci. U.S.A.}\ }\textbf {\bibinfo
  {volume} {101}},\ \bibinfo {pages} {5743} (\bibinfo {year}
  {2004})}\BibitemShut {NoStop}%
\bibitem [{\citenamefont {Khavrutskii}, \citenamefont {Musaev},\ and\
  \citenamefont {Morokuma}(2005)}]{KhavrutskiiDFT5}%
  \BibitemOpen
  \bibfield  {author} {\bibinfo {author} {\bibfnamefont {I.~V.}\ \bibnamefont
  {Khavrutskii}}, \bibinfo {author} {\bibfnamefont {D.~G.}\ \bibnamefont
  {Musaev}}, \ and\ \bibinfo {author} {\bibfnamefont {K.}~\bibnamefont
  {Morokuma}},\ }\href {\doibase 10.1021/ic0490122} {\bibfield  {journal}
  {\bibinfo  {journal} {Inorg. Chem.}\ }\textbf {\bibinfo {volume} {44}},\
  \bibinfo {pages} {306} (\bibinfo {year} {2005})}\BibitemShut {NoStop}%
\bibitem [{\citenamefont {Khavrutskii}\ \emph {et~al.}(2004)\citenamefont
  {Khavrutskii}, \citenamefont {Rahim}, \citenamefont {Musaev},\ and\
  \citenamefont {Morokuma}}]{KhavrutskiiDFT4}%
  \BibitemOpen
  \bibfield  {author} {\bibinfo {author} {\bibfnamefont {I.~V.}\ \bibnamefont
  {Khavrutskii}}, \bibinfo {author} {\bibfnamefont {R.~R.}\ \bibnamefont
  {Rahim}}, \bibinfo {author} {\bibfnamefont {D.~G.}\ \bibnamefont {Musaev}}, \
  and\ \bibinfo {author} {\bibfnamefont {K.}~\bibnamefont {Morokuma}},\ }\href
  {\doibase 10.1021/jp0496912} {\bibfield  {journal} {\bibinfo  {journal} {J.
  Phys. Chem. B}\ }\textbf {\bibinfo {volume} {108}},\ \bibinfo {pages} {3845}
  (\bibinfo {year} {2004})}\BibitemShut {NoStop}%
\bibitem [{\citenamefont {Abashkin}\ and\ \citenamefont
  {Burt}(2004)}]{AbashkinDFT2}%
  \BibitemOpen
  \bibfield  {author} {\bibinfo {author} {\bibfnamefont {Y.~G.}\ \bibnamefont
  {Abashkin}}\ and\ \bibinfo {author} {\bibfnamefont {S.~K.}\ \bibnamefont
  {Burt}},\ }\href {\doibase 10.1021/ol036051t} {\bibfield  {journal} {\bibinfo
   {journal} {Org. Lett.}\ }\textbf {\bibinfo {volume} {6}},\ \bibinfo {pages}
  {59} (\bibinfo {year} {2004})}\BibitemShut {NoStop}%
\bibitem [{\citenamefont {Scheurer}\ \emph {et~al.}(2005)\citenamefont
  {Scheurer}, \citenamefont {Maid}, \citenamefont {Hampel}, \citenamefont
  {Saalfrank}, \citenamefont {Toupet}, \citenamefont {Mosset}, \citenamefont
  {Puchta},\ and\ \citenamefont {van Eikema~Hommes}}]{EJOC200500042}%
  \BibitemOpen
  \bibfield  {author} {\bibinfo {author} {\bibfnamefont {A.}~\bibnamefont
  {Scheurer}}, \bibinfo {author} {\bibfnamefont {H.}~\bibnamefont {Maid}},
  \bibinfo {author} {\bibfnamefont {F.}~\bibnamefont {Hampel}}, \bibinfo
  {author} {\bibfnamefont {R.~W.}\ \bibnamefont {Saalfrank}}, \bibinfo {author}
  {\bibfnamefont {L.}~\bibnamefont {Toupet}}, \bibinfo {author} {\bibfnamefont
  {P.}~\bibnamefont {Mosset}}, \bibinfo {author} {\bibfnamefont
  {R.}~\bibnamefont {Puchta}}, \ and\ \bibinfo {author} {\bibfnamefont
  {N.~J.~R.}\ \bibnamefont {van Eikema~Hommes}},\ }\href {\doibase
  10.1002/ejoc.200500042} {\bibfield  {journal} {\bibinfo  {journal} {Eur. J.
  Org. Chem.}\ }\textbf {\bibinfo {volume} {2005}},\ \bibinfo {pages} {2566}
  (\bibinfo {year} {2005})}\BibitemShut {NoStop}%
\bibitem [{\citenamefont {Abashkin}, \citenamefont {Collins},\ and\
  \citenamefont {Burt}(2001)}]{AbashkinDFT1}%
  \BibitemOpen
  \bibfield  {author} {\bibinfo {author} {\bibfnamefont {Y.~G.}\ \bibnamefont
  {Abashkin}}, \bibinfo {author} {\bibfnamefont {J.~R.}\ \bibnamefont
  {Collins}}, \ and\ \bibinfo {author} {\bibfnamefont {S.~K.}\ \bibnamefont
  {Burt}},\ }\href {\doibase 10.1021/ic0012221} {\bibfield  {journal} {\bibinfo
   {journal} {Inorg. Chem.}\ }\textbf {\bibinfo {volume} {40}},\ \bibinfo
  {pages} {4040} (\bibinfo {year} {2001})}\BibitemShut {NoStop}%
\bibitem [{\citenamefont {Sears}\ and\ \citenamefont
  {Sherrill}(2008)}]{sears2}%
  \BibitemOpen
  \bibfield  {author} {\bibinfo {author} {\bibfnamefont {J.~S.}\ \bibnamefont
  {Sears}}\ and\ \bibinfo {author} {\bibfnamefont {C.~D.}\ \bibnamefont
  {Sherrill}},\ }\href {\doibase 10.1021/jp711595w} {\bibfield  {journal}
  {\bibinfo  {journal} {J. Phys. Chem. A}\ }\textbf {\bibinfo {volume} {112}},\
  \bibinfo {pages} {3466} (\bibinfo {year} {2008})}\BibitemShut {NoStop}%
\bibitem [{\citenamefont {Takatani}, \citenamefont {Sears},\ and\ \citenamefont
  {Sherrill}(2010)}]{sears3}%
  \BibitemOpen
  \bibfield  {author} {\bibinfo {author} {\bibfnamefont {T.}~\bibnamefont
  {Takatani}}, \bibinfo {author} {\bibfnamefont {J.~S.}\ \bibnamefont {Sears}},
  \ and\ \bibinfo {author} {\bibfnamefont {C.~D.}\ \bibnamefont {Sherrill}},\
  }\href {\doibase 10.1021/jp1046084} {\bibfield  {journal} {\bibinfo
  {journal} {J. Phys. Chem. A}\ }\textbf {\bibinfo {volume} {114}},\ \bibinfo
  {pages} {11714} (\bibinfo {year} {2010})}\BibitemShut {NoStop}%
\bibitem [{\citenamefont {Ivanic}, \citenamefont {Collins},\ and\ \citenamefont
  {Burt}(2004)}]{Ivanic2}%
  \BibitemOpen
  \bibfield  {author} {\bibinfo {author} {\bibfnamefont {J.}~\bibnamefont
  {Ivanic}}, \bibinfo {author} {\bibfnamefont {J.~R.}\ \bibnamefont {Collins}},
  \ and\ \bibinfo {author} {\bibfnamefont {S.~K.}\ \bibnamefont {Burt}},\
  }\href {\doibase 10.1021/jp031214g} {\bibfield  {journal} {\bibinfo
  {journal} {J. Phys. Chem. A}\ }\textbf {\bibinfo {volume} {108}},\ \bibinfo
  {pages} {2314} (\bibinfo {year} {2004})}\BibitemShut {NoStop}%
\bibitem [{\citenamefont {Sears}\ and\ \citenamefont {Sherrill}(2006)}]{Sears}%
  \BibitemOpen
  \bibfield  {author} {\bibinfo {author} {\bibfnamefont {J.~S.}\ \bibnamefont
  {Sears}}\ and\ \bibinfo {author} {\bibfnamefont {C.~D.}\ \bibnamefont
  {Sherrill}},\ }\href {\doibase 10.1063/1.2187974} {\bibfield  {journal}
  {\bibinfo  {journal} {J. Chem. Phys.}\ }\textbf {\bibinfo {volume} {124}},\
  \bibinfo {eid} {144314} (\bibinfo {year} {2006})}\BibitemShut {NoStop}%
\bibitem [{\citenamefont {Ma}, \citenamefont {Li~Manni},\ and\ \citenamefont
  {Gagliardi}(2011)}]{Gagliardi}%
  \BibitemOpen
  \bibfield  {author} {\bibinfo {author} {\bibfnamefont {D.}~\bibnamefont
  {Ma}}, \bibinfo {author} {\bibfnamefont {G.}~\bibnamefont {Li~Manni}}, \ and\
  \bibinfo {author} {\bibfnamefont {L.}~\bibnamefont {Gagliardi}},\ }\href
  {\doibase 10.1063/1.3611401} {\bibfield  {journal} {\bibinfo  {journal} {J.
  Chem. Phys.}\ }\textbf {\bibinfo {volume} {135}},\ \bibinfo {eid} {044128}
  (\bibinfo {year} {2011})}\BibitemShut {NoStop}%
\bibitem [{\citenamefont {Hariharan}\ and\ \citenamefont
  {Pople}(1973)}]{631gstar1}%
  \BibitemOpen
  \bibfield  {author} {\bibinfo {author} {\bibfnamefont {P.~C.}\ \bibnamefont
  {Hariharan}}\ and\ \bibinfo {author} {\bibfnamefont {J.~A.}\ \bibnamefont
  {Pople}},\ }\href {\doibase 10.1007/BF00533485} {\bibfield  {journal}
  {\bibinfo  {journal} {Theor. Chim. Acta}\ }\textbf {\bibinfo {volume} {28}},\
  \bibinfo {pages} {213} (\bibinfo {year} {1973})}\BibitemShut {NoStop}%
\bibitem [{\citenamefont {Dunning}(1989)}]{ccpvdzbasis}%
  \BibitemOpen
  \bibfield  {author} {\bibinfo {author} {\bibfnamefont {T.~H.}\ \bibnamefont
  {Dunning}},\ }\href {\doibase 10.1063/1.456153} {\bibfield  {journal}
  {\bibinfo  {journal} {J. Chem. Phys.}\ }\textbf {\bibinfo {volume} {90}},\
  \bibinfo {pages} {1007} (\bibinfo {year} {1989})}\BibitemShut {NoStop}%
\bibitem [{\citenamefont {Widmark}, \citenamefont {Malmqvist},\ and\
  \citenamefont {Roos}(1990)}]{anorccvdzp}%
  \BibitemOpen
  \bibfield  {author} {\bibinfo {author} {\bibfnamefont {P.-O.}\ \bibnamefont
  {Widmark}}, \bibinfo {author} {\bibfnamefont {P.-A.}\ \bibnamefont
  {Malmqvist}}, \ and\ \bibinfo {author} {\bibfnamefont {B.~O.}\ \bibnamefont
  {Roos}},\ }\href {\doibase 10.1007/BF01120130} {\bibfield  {journal}
  {\bibinfo  {journal} {Theor. Chim. Acta}\ }\textbf {\bibinfo {volume} {77}},\
  \bibinfo {pages} {291} (\bibinfo {year} {1990})}\BibitemShut {NoStop}%
\bibitem [{\citenamefont {Reiher}\ and\ \citenamefont
  {Wolf}(2004)}]{DKH2citation}%
  \BibitemOpen
  \bibfield  {author} {\bibinfo {author} {\bibfnamefont {M.}~\bibnamefont
  {Reiher}}\ and\ \bibinfo {author} {\bibfnamefont {A.}~\bibnamefont {Wolf}},\
  }\href {\doibase 10.1063/1.1768160} {\bibfield  {journal} {\bibinfo
  {journal} {J. Chem. Phys.}\ }\textbf {\bibinfo {volume} {121}},\ \bibinfo
  {pages} {2037} (\bibinfo {year} {2004})}\BibitemShut {NoStop}%
\bibitem [{\citenamefont {Turney}\ \emph {et~al.}(2012)\citenamefont {Turney},
  \citenamefont {Simmonett}, \citenamefont {Parrish}, \citenamefont
  {Hohenstein}, \citenamefont {Evangelista}, \citenamefont {Fermann},
  \citenamefont {Mintz}, \citenamefont {Burns}, \citenamefont {Wilke},
  \citenamefont {Abrams}, \citenamefont {Russ}, \citenamefont {Leininger},
  \citenamefont {Janssen}, \citenamefont {Seidl}, \citenamefont {Allen},
  \citenamefont {Schaefer}, \citenamefont {King}, \citenamefont {Valeev},
  \citenamefont {Sherrill},\ and\ \citenamefont {Crawford}}]{Psi4}%
  \BibitemOpen
  \bibfield  {author} {\bibinfo {author} {\bibfnamefont {J.~M.}\ \bibnamefont
  {Turney}}, \bibinfo {author} {\bibfnamefont {A.~C.}\ \bibnamefont
  {Simmonett}}, \bibinfo {author} {\bibfnamefont {R.~M.}\ \bibnamefont
  {Parrish}}, \bibinfo {author} {\bibfnamefont {E.~G.}\ \bibnamefont
  {Hohenstein}}, \bibinfo {author} {\bibfnamefont {F.~A.}\ \bibnamefont
  {Evangelista}}, \bibinfo {author} {\bibfnamefont {J.~T.}\ \bibnamefont
  {Fermann}}, \bibinfo {author} {\bibfnamefont {B.~J.}\ \bibnamefont {Mintz}},
  \bibinfo {author} {\bibfnamefont {L.~A.}\ \bibnamefont {Burns}}, \bibinfo
  {author} {\bibfnamefont {J.~J.}\ \bibnamefont {Wilke}}, \bibinfo {author}
  {\bibfnamefont {M.~L.}\ \bibnamefont {Abrams}}, \bibinfo {author}
  {\bibfnamefont {N.~J.}\ \bibnamefont {Russ}}, \bibinfo {author}
  {\bibfnamefont {M.~L.}\ \bibnamefont {Leininger}}, \bibinfo {author}
  {\bibfnamefont {C.~L.}\ \bibnamefont {Janssen}}, \bibinfo {author}
  {\bibfnamefont {E.~T.}\ \bibnamefont {Seidl}}, \bibinfo {author}
  {\bibfnamefont {W.~D.}\ \bibnamefont {Allen}}, \bibinfo {author}
  {\bibfnamefont {H.~F.}\ \bibnamefont {Schaefer}}, \bibinfo {author}
  {\bibfnamefont {R.~A.}\ \bibnamefont {King}}, \bibinfo {author}
  {\bibfnamefont {E.~F.}\ \bibnamefont {Valeev}}, \bibinfo {author}
  {\bibfnamefont {C.~D.}\ \bibnamefont {Sherrill}}, \ and\ \bibinfo {author}
  {\bibfnamefont {T.~D.}\ \bibnamefont {Crawford}},\ }\href {\doibase
  10.1002/wcms.93} {\bibfield  {journal} {\bibinfo  {journal} {WIREs Comput.
  Mol. Sci.}\ }\textbf {\bibinfo {volume} {2}},\ \bibinfo {pages} {556}
  (\bibinfo {year} {2012})}\BibitemShut {NoStop}%
\bibitem [{\citenamefont {Werner}\ \emph {et~al.}(2012)\citenamefont {Werner},
  \citenamefont {Knowles}, \citenamefont {Knizia}, \citenamefont {Manby},\ and\
  \citenamefont {Sch{\"u}tz}}]{MOLPRO-WIREs}%
  \BibitemOpen
  \bibfield  {author} {\bibinfo {author} {\bibfnamefont {H.-J.}\ \bibnamefont
  {Werner}}, \bibinfo {author} {\bibfnamefont {P.~J.}\ \bibnamefont {Knowles}},
  \bibinfo {author} {\bibfnamefont {G.}~\bibnamefont {Knizia}}, \bibinfo
  {author} {\bibfnamefont {F.~R.}\ \bibnamefont {Manby}}, \ and\ \bibinfo
  {author} {\bibfnamefont {M.}~\bibnamefont {Sch{\"u}tz}},\ }\href {\doibase
  10.1002/wcms.82} {\bibfield  {journal} {\bibinfo  {journal} {WIREs Comput.
  Mol. Sci.}\ }\textbf {\bibinfo {volume} {2}},\ \bibinfo {pages} {242}
  (\bibinfo {year} {2012})}\BibitemShut {NoStop}%
\bibitem [{\citenamefont {Chan}\ and\ \citenamefont
  {Head-Gordon}(2002)}]{ChanQCDMRG}%
  \BibitemOpen
  \bibfield  {author} {\bibinfo {author} {\bibfnamefont {G.~K.-L.}\
  \bibnamefont {Chan}}\ and\ \bibinfo {author} {\bibfnamefont {M.}~\bibnamefont
  {Head-Gordon}},\ }\href {\doibase 10.1063/1.1449459} {\bibfield  {journal}
  {\bibinfo  {journal} {J. Chem. Phys.}\ }\textbf {\bibinfo {volume} {116}},\
  \bibinfo {pages} {4462} (\bibinfo {year} {2002})}\BibitemShut {NoStop}%
\bibitem [{\citenamefont {{Wouters}}\ \emph {et~al.}(2014)\citenamefont
  {{Wouters}}, \citenamefont {{Poelmans}}, \citenamefont {{Ayers}},\ and\
  \citenamefont {{Van Neck}}}]{CheMPS2paper}%
  \BibitemOpen
  \bibfield  {author} {\bibinfo {author} {\bibfnamefont {S.}~\bibnamefont
  {{Wouters}}}, \bibinfo {author} {\bibfnamefont {W.}~\bibnamefont
  {{Poelmans}}}, \bibinfo {author} {\bibfnamefont {P.~W.}\ \bibnamefont
  {{Ayers}}}, \ and\ \bibinfo {author} {\bibfnamefont {D.}~\bibnamefont {{Van
  Neck}}},\ }\href {\doibase 10.1016/j.cpc.2014.01.019} {\bibfield  {journal}
  {\bibinfo  {journal} {Comput. Phys. Commun.}\ }\textbf {\bibinfo {volume}
  {185}},\ \bibinfo {pages} {1501} (\bibinfo {year} {2014})}\BibitemShut
  {NoStop}%
\bibitem [{\citenamefont {White}\ and\ \citenamefont
  {Martin}(1999)}]{WhiteQCDMRG}%
  \BibitemOpen
  \bibfield  {author} {\bibinfo {author} {\bibfnamefont {S.~R.}\ \bibnamefont
  {White}}\ and\ \bibinfo {author} {\bibfnamefont {R.~L.}\ \bibnamefont
  {Martin}},\ }\href {\doibase 10.1063/1.478295} {\bibfield  {journal}
  {\bibinfo  {journal} {J. Chem. Phys.}\ }\textbf {\bibinfo {volume} {110}},\
  \bibinfo {pages} {4127} (\bibinfo {year} {1999})}\BibitemShut {NoStop}%
\bibitem [{\citenamefont {Mitrushenkov}\ \emph {et~al.}(2001)\citenamefont
  {Mitrushenkov}, \citenamefont {Fano}, \citenamefont {Ortolani}, \citenamefont
  {Linguerri},\ and\ \citenamefont {Palmieri}}]{Mitrushenkov}%
  \BibitemOpen
  \bibfield  {author} {\bibinfo {author} {\bibfnamefont {A.~O.}\ \bibnamefont
  {Mitrushenkov}}, \bibinfo {author} {\bibfnamefont {G.}~\bibnamefont {Fano}},
  \bibinfo {author} {\bibfnamefont {F.}~\bibnamefont {Ortolani}}, \bibinfo
  {author} {\bibfnamefont {R.}~\bibnamefont {Linguerri}}, \ and\ \bibinfo
  {author} {\bibfnamefont {P.}~\bibnamefont {Palmieri}},\ }\href {\doibase
  10.1063/1.1389475} {\bibfield  {journal} {\bibinfo  {journal} {J. Chem.
  Phys.}\ }\textbf {\bibinfo {volume} {115}},\ \bibinfo {pages} {6815}
  (\bibinfo {year} {2001})}\BibitemShut {NoStop}%
\bibitem [{\citenamefont {Legeza}, \citenamefont {R\"oder},\ and\ \citenamefont
  {Hess}(2003)}]{PhysRevB.67.125114}%
  \BibitemOpen
  \bibfield  {author} {\bibinfo {author} {\bibfnamefont {O.}~\bibnamefont
  {Legeza}}, \bibinfo {author} {\bibfnamefont {J.}~\bibnamefont {R\"oder}}, \
  and\ \bibinfo {author} {\bibfnamefont {B.~A.}\ \bibnamefont {Hess}},\ }\href
  {\doibase 10.1103/PhysRevB.67.125114} {\bibfield  {journal} {\bibinfo
  {journal} {Phys. Rev. B}\ }\textbf {\bibinfo {volume} {67}},\ \bibinfo
  {pages} {125114} (\bibinfo {year} {2003})}\BibitemShut {NoStop}%
\bibitem [{\citenamefont {Moritz}, \citenamefont {Hess},\ and\ \citenamefont
  {Reiher}(2005)}]{ReiherCode}%
  \BibitemOpen
  \bibfield  {author} {\bibinfo {author} {\bibfnamefont {G.}~\bibnamefont
  {Moritz}}, \bibinfo {author} {\bibfnamefont {B.~A.}\ \bibnamefont {Hess}}, \
  and\ \bibinfo {author} {\bibfnamefont {M.}~\bibnamefont {Reiher}},\ }\href
  {\doibase 10.1063/1.1824891} {\bibfield  {journal} {\bibinfo  {journal} {J.
  Chem. Phys.}\ }\textbf {\bibinfo {volume} {122}},\ \bibinfo {pages} {024107}
  (\bibinfo {year} {2005})}\BibitemShut {NoStop}%
\bibitem [{\citenamefont {Zgid}\ and\ \citenamefont
  {Nooijen}(2008{\natexlab{a}})}]{Zgid}%
  \BibitemOpen
  \bibfield  {author} {\bibinfo {author} {\bibfnamefont {D.}~\bibnamefont
  {Zgid}}\ and\ \bibinfo {author} {\bibfnamefont {M.}~\bibnamefont {Nooijen}},\
  }\href {\doibase 10.1063/1.2814150} {\bibfield  {journal} {\bibinfo
  {journal} {J. Chem. Phys.}\ }\textbf {\bibinfo {volume} {128}},\ \bibinfo
  {pages} {014107} (\bibinfo {year} {2008}{\natexlab{a}})}\BibitemShut
  {NoStop}%
\bibitem [{\citenamefont {Kurashige}\ and\ \citenamefont
  {Yanai}(2009)}]{KurashigeBasic}%
  \BibitemOpen
  \bibfield  {author} {\bibinfo {author} {\bibfnamefont {Y.}~\bibnamefont
  {Kurashige}}\ and\ \bibinfo {author} {\bibfnamefont {T.}~\bibnamefont
  {Yanai}},\ }\href {\doibase 10.1063/1.3152576} {\bibfield  {journal}
  {\bibinfo  {journal} {J. Chem. Phys.}\ }\textbf {\bibinfo {volume} {130}},\
  \bibinfo {pages} {234114} (\bibinfo {year} {2009})}\BibitemShut {NoStop}%
\bibitem [{\citenamefont {Luo}, \citenamefont {Qin},\ and\ \citenamefont
  {Xiang}(2010)}]{PhysRevB.81.235129}%
  \BibitemOpen
  \bibfield  {author} {\bibinfo {author} {\bibfnamefont {H.-G.}\ \bibnamefont
  {Luo}}, \bibinfo {author} {\bibfnamefont {M.-P.}\ \bibnamefont {Qin}}, \ and\
  \bibinfo {author} {\bibfnamefont {T.}~\bibnamefont {Xiang}},\ }\href
  {\doibase 10.1103/PhysRevB.81.235129} {\bibfield  {journal} {\bibinfo
  {journal} {Phys. Rev. B}\ }\textbf {\bibinfo {volume} {81}},\ \bibinfo
  {pages} {235129} (\bibinfo {year} {2010})}\BibitemShut {NoStop}%
\bibitem [{\citenamefont {Sharma}\ and\ \citenamefont
  {Chan}(2012)}]{Sharma2012}%
  \BibitemOpen
  \bibfield  {author} {\bibinfo {author} {\bibfnamefont {S.}~\bibnamefont
  {Sharma}}\ and\ \bibinfo {author} {\bibfnamefont {G.~K.-L.}\ \bibnamefont
  {Chan}},\ }\href {\doibase 10.1063/1.3695642} {\bibfield  {journal} {\bibinfo
   {journal} {J. Chem. Phys.}\ }\textbf {\bibinfo {volume} {136}},\ \bibinfo
  {pages} {124121} (\bibinfo {year} {2012})}\BibitemShut {NoStop}%
\bibitem [{\citenamefont {Wouters}\ \emph {et~al.}(2012)\citenamefont
  {Wouters}, \citenamefont {Limacher}, \citenamefont {{Van Neck}},\ and\
  \citenamefont {Ayers}}]{wouters2012}%
  \BibitemOpen
  \bibfield  {author} {\bibinfo {author} {\bibfnamefont {S.}~\bibnamefont
  {Wouters}}, \bibinfo {author} {\bibfnamefont {P.~A.}\ \bibnamefont
  {Limacher}}, \bibinfo {author} {\bibfnamefont {D.}~\bibnamefont {{Van
  Neck}}}, \ and\ \bibinfo {author} {\bibfnamefont {P.~W.}\ \bibnamefont
  {Ayers}},\ }\href {\doibase 10.1063/1.3700087} {\bibfield  {journal}
  {\bibinfo  {journal} {J. Chem. Phys.}\ }\textbf {\bibinfo {volume} {136}},\
  \bibinfo {pages} {134110} (\bibinfo {year} {2012})}\BibitemShut {NoStop}%
\bibitem [{\citenamefont {Legeza}\ and\ \citenamefont
  {F\'ath}(1996)}]{PhysRevB.53.14349}%
  \BibitemOpen
  \bibfield  {author} {\bibinfo {author} {\bibfnamefont {O.}~\bibnamefont
  {Legeza}}\ and\ \bibinfo {author} {\bibfnamefont {G.}~\bibnamefont
  {F\'ath}},\ }\href {\doibase 10.1103/PhysRevB.53.14349} {\bibfield  {journal}
  {\bibinfo  {journal} {Phys. Rev. B}\ }\textbf {\bibinfo {volume} {53}},\
  \bibinfo {pages} {14349} (\bibinfo {year} {1996})}\BibitemShut {NoStop}%
\bibitem [{\citenamefont {Wouters}(2014)}]{CheMPS2github}%
  \BibitemOpen
  \bibfield  {author} {\bibinfo {author} {\bibfnamefont {S.}~\bibnamefont
  {Wouters}},\ }\href@noop {} {\enquote {\bibinfo {title} {{CheMPS2}: a
  spin-adapted implementation of {DMRG} for ab initio quantum chemistry},}\
  }\bibinfo {howpublished} {\url{https://github.com/SebWouters/CheMPS2}}
  (\bibinfo {year} {2014})\BibitemShut {NoStop}%
\bibitem [{Note1()}]{Note1}%
  \BibitemOpen
  \bibinfo {note} {Each spin multiplet in a spin-adapted MPS is represented by
  only one reduced basis state. This is the \protect \textit {reduced} to which
  is referred in the caption of Fig. \ref {FIG-Extrapol}.}\BibitemShut {Stop}%
\bibitem [{\citenamefont {Zgid}\ and\ \citenamefont
  {Nooijen}(2008{\natexlab{b}})}]{Zgid-2RDM}%
  \BibitemOpen
  \bibfield  {author} {\bibinfo {author} {\bibfnamefont {D.}~\bibnamefont
  {Zgid}}\ and\ \bibinfo {author} {\bibfnamefont {M.}~\bibnamefont {Nooijen}},\
  }\href {\doibase 10.1063/1.2883980} {\bibfield  {journal} {\bibinfo
  {journal} {J. Chem. Phys.}\ }\textbf {\bibinfo {volume} {128}},\ \bibinfo
  {pages} {144115} (\bibinfo {year} {2008}{\natexlab{b}})}\BibitemShut
  {NoStop}%
\bibitem [{\citenamefont {Ghosh}\ \emph {et~al.}(2008)\citenamefont {Ghosh},
  \citenamefont {Hachmann}, \citenamefont {Yanai},\ and\ \citenamefont
  {Chan}}]{Ghosh-DMRGSCF}%
  \BibitemOpen
  \bibfield  {author} {\bibinfo {author} {\bibfnamefont {D.}~\bibnamefont
  {Ghosh}}, \bibinfo {author} {\bibfnamefont {J.}~\bibnamefont {Hachmann}},
  \bibinfo {author} {\bibfnamefont {T.}~\bibnamefont {Yanai}}, \ and\ \bibinfo
  {author} {\bibfnamefont {G.~K.-L.}\ \bibnamefont {Chan}},\ }\href {\doibase
  10.1063/1.2883976} {\bibfield  {journal} {\bibinfo  {journal} {J. Chem.
  Phys.}\ }\textbf {\bibinfo {volume} {128}},\ \bibinfo {pages} {144117}
  (\bibinfo {year} {2008})}\BibitemShut {NoStop}%
\bibitem [{\citenamefont {Siegbahn}\ \emph {et~al.}(1981)\citenamefont
  {Siegbahn}, \citenamefont {Alml\"of}, \citenamefont {Heiberg},\ and\
  \citenamefont {Roos}}]{RDMbasedCASSCF}%
  \BibitemOpen
  \bibfield  {author} {\bibinfo {author} {\bibfnamefont {P.~E.~M.}\
  \bibnamefont {Siegbahn}}, \bibinfo {author} {\bibfnamefont {J.}~\bibnamefont
  {Alml\"of}}, \bibinfo {author} {\bibfnamefont {A.}~\bibnamefont {Heiberg}}, \
  and\ \bibinfo {author} {\bibfnamefont {B.~O.}\ \bibnamefont {Roos}},\ }\href
  {\doibase 10.1063/1.441359} {\bibfield  {journal} {\bibinfo  {journal} {J.
  Chem. Phys.}\ }\textbf {\bibinfo {volume} {74}},\ \bibinfo {pages} {2384}
  (\bibinfo {year} {1981})}\BibitemShut {NoStop}%
\bibitem [{\citenamefont {Zgid}\ and\ \citenamefont
  {Nooijen}(2008{\natexlab{c}})}]{Zgid-DMRGSCF}%
  \BibitemOpen
  \bibfield  {author} {\bibinfo {author} {\bibfnamefont {D.}~\bibnamefont
  {Zgid}}\ and\ \bibinfo {author} {\bibfnamefont {M.}~\bibnamefont {Nooijen}},\
  }\href {\doibase 10.1063/1.2883981} {\bibfield  {journal} {\bibinfo
  {journal} {J. Chem. Phys.}\ }\textbf {\bibinfo {volume} {128}},\ \bibinfo
  {pages} {144116} (\bibinfo {year} {2008}{\natexlab{c}})}\BibitemShut
  {NoStop}%
\bibitem [{\citenamefont {Yanai}\ \emph {et~al.}(2009)\citenamefont {Yanai},
  \citenamefont {Kurashige}, \citenamefont {Ghosh},\ and\ \citenamefont
  {Chan}}]{QUA22099}%
  \BibitemOpen
  \bibfield  {author} {\bibinfo {author} {\bibfnamefont {T.}~\bibnamefont
  {Yanai}}, \bibinfo {author} {\bibfnamefont {Y.}~\bibnamefont {Kurashige}},
  \bibinfo {author} {\bibfnamefont {D.}~\bibnamefont {Ghosh}}, \ and\ \bibinfo
  {author} {\bibfnamefont {G.~K.-L.}\ \bibnamefont {Chan}},\ }\href {\doibase
  10.1002/qua.22099} {\bibfield  {journal} {\bibinfo  {journal} {Int. J. Quant.
  Chem.}\ }\textbf {\bibinfo {volume} {109}},\ \bibinfo {pages} {2178}
  (\bibinfo {year} {2009})}\BibitemShut {NoStop}%
\bibitem [{\citenamefont {Moritz}\ and\ \citenamefont
  {Reiher}(2006)}]{Reiher2}%
  \BibitemOpen
  \bibfield  {author} {\bibinfo {author} {\bibfnamefont {G.}~\bibnamefont
  {Moritz}}\ and\ \bibinfo {author} {\bibfnamefont {M.}~\bibnamefont
  {Reiher}},\ }\href {\doibase 10.1063/1.2139998} {\bibfield  {journal}
  {\bibinfo  {journal} {J. Chem. Phys.}\ }\textbf {\bibinfo {volume} {124}},\
  \bibinfo {pages} {034103} (\bibinfo {year} {2006})}\BibitemShut {NoStop}%
\bibitem [{\citenamefont {Marti}\ \emph {et~al.}(2008)\citenamefont {Marti},
  \citenamefont {Ondík}, \citenamefont {Moritz},\ and\ \citenamefont
  {Reiher}}]{Reiher3}%
  \BibitemOpen
  \bibfield  {author} {\bibinfo {author} {\bibfnamefont {K.~H.}\ \bibnamefont
  {Marti}}, \bibinfo {author} {\bibfnamefont {I.~M.}\ \bibnamefont {Ondík}},
  \bibinfo {author} {\bibfnamefont {G.}~\bibnamefont {Moritz}}, \ and\ \bibinfo
  {author} {\bibfnamefont {M.}~\bibnamefont {Reiher}},\ }\href {\doibase
  10.1063/1.2805383} {\bibfield  {journal} {\bibinfo  {journal} {J. Chem.
  Phys.}\ }\textbf {\bibinfo {volume} {128}},\ \bibinfo {pages} {014104}
  (\bibinfo {year} {2008})}\BibitemShut {NoStop}%
\bibitem [{\citenamefont {Yanai}\ \emph {et~al.}(2010)\citenamefont {Yanai},
  \citenamefont {Kurashige}, \citenamefont {Neuscamman},\ and\ \citenamefont
  {Chan}}]{DMRG-CT}%
  \BibitemOpen
  \bibfield  {author} {\bibinfo {author} {\bibfnamefont {T.}~\bibnamefont
  {Yanai}}, \bibinfo {author} {\bibfnamefont {Y.}~\bibnamefont {Kurashige}},
  \bibinfo {author} {\bibfnamefont {E.}~\bibnamefont {Neuscamman}}, \ and\
  \bibinfo {author} {\bibfnamefont {G.~K.-L.}\ \bibnamefont {Chan}},\ }\href
  {\doibase 10.1063/1.3275806} {\bibfield  {journal} {\bibinfo  {journal} {J.
  Chem. Phys.}\ }\textbf {\bibinfo {volume} {132}},\ \bibinfo {pages} {024105}
  (\bibinfo {year} {2010})}\BibitemShut {NoStop}%
\bibitem [{\citenamefont {Kurashige}\ and\ \citenamefont
  {Yanai}(2011)}]{DMRG-CASPT2}%
  \BibitemOpen
  \bibfield  {author} {\bibinfo {author} {\bibfnamefont {Y.}~\bibnamefont
  {Kurashige}}\ and\ \bibinfo {author} {\bibfnamefont {T.}~\bibnamefont
  {Yanai}},\ }\href {\doibase 10.1063/1.3629454} {\bibfield  {journal}
  {\bibinfo  {journal} {J. Chem. Phys.}\ }\textbf {\bibinfo {volume} {135}},\
  \bibinfo {pages} {094104} (\bibinfo {year} {2011})}\BibitemShut {NoStop}%
\bibitem [{\citenamefont {Boguslawski}\ \emph
  {et~al.}(2012{\natexlab{a}})\citenamefont {Boguslawski}, \citenamefont
  {Marti}, \citenamefont {Legeza},\ and\ \citenamefont {Reiher}}]{ct300211j}%
  \BibitemOpen
  \bibfield  {author} {\bibinfo {author} {\bibfnamefont {K.}~\bibnamefont
  {Boguslawski}}, \bibinfo {author} {\bibfnamefont {K.~H.}\ \bibnamefont
  {Marti}}, \bibinfo {author} {\bibfnamefont {Ã.}~\bibnamefont {Legeza}}, \
  and\ \bibinfo {author} {\bibfnamefont {M.}~\bibnamefont {Reiher}},\ }\href
  {\doibase 10.1021/ct300211j} {\bibfield  {journal} {\bibinfo  {journal} {J.
  Chem. Theory Comput.}\ }\textbf {\bibinfo {volume} {8}},\ \bibinfo {pages}
  {1970} (\bibinfo {year} {2012}{\natexlab{a}})}\BibitemShut {NoStop}%
\bibitem [{\citenamefont {Boguslawski}\ \emph
  {et~al.}(2012{\natexlab{b}})\citenamefont {Boguslawski}, \citenamefont
  {Tecmer}, \citenamefont {Legeza},\ and\ \citenamefont {Reiher}}]{jz301319v}%
  \BibitemOpen
  \bibfield  {author} {\bibinfo {author} {\bibfnamefont {K.}~\bibnamefont
  {Boguslawski}}, \bibinfo {author} {\bibfnamefont {P.}~\bibnamefont {Tecmer}},
  \bibinfo {author} {\bibfnamefont {O.}~\bibnamefont {Legeza}}, \ and\ \bibinfo
  {author} {\bibfnamefont {M.}~\bibnamefont {Reiher}},\ }\href {\doibase
  10.1021/jz301319v} {\bibfield  {journal} {\bibinfo  {journal} {J. Phys. Chem.
  Lett.}\ }\textbf {\bibinfo {volume} {3}},\ \bibinfo {pages} {3129} (\bibinfo
  {year} {2012}{\natexlab{b}})}\BibitemShut {NoStop}%
\bibitem [{\citenamefont {Kurashige}, \citenamefont {Chan},\ and\ \citenamefont
  {Yanai}(2013)}]{naturechem}%
  \BibitemOpen
  \bibfield  {author} {\bibinfo {author} {\bibfnamefont {Y.}~\bibnamefont
  {Kurashige}}, \bibinfo {author} {\bibfnamefont {G.~K.-L.}\ \bibnamefont
  {Chan}}, \ and\ \bibinfo {author} {\bibfnamefont {T.}~\bibnamefont {Yanai}},\
  }\href {\doibase 10.1038/nchem.1677} {\bibfield  {journal} {\bibinfo
  {journal} {Nat. Chem.}\ }\textbf {\bibinfo {volume} {5}},\ \bibinfo {pages}
  {660} (\bibinfo {year} {2013})}\BibitemShut {NoStop}%
\bibitem [{\citenamefont {Harris}\ \emph {et~al.}(2014)\citenamefont {Harris},
  \citenamefont {Kurashige}, \citenamefont {Yanai},\ and\ \citenamefont
  {Morokuma}}]{1.4863345}%
  \BibitemOpen
  \bibfield  {author} {\bibinfo {author} {\bibfnamefont {T.~V.}\ \bibnamefont
  {Harris}}, \bibinfo {author} {\bibfnamefont {Y.}~\bibnamefont {Kurashige}},
  \bibinfo {author} {\bibfnamefont {T.}~\bibnamefont {Yanai}}, \ and\ \bibinfo
  {author} {\bibfnamefont {K.}~\bibnamefont {Morokuma}},\ }\href {\doibase
  10.1063/1.4863345} {\bibfield  {journal} {\bibinfo  {journal} {J. Chem.
  Phys.}\ }\textbf {\bibinfo {volume} {140}},\ \bibinfo {eid} {054303}
  (\bibinfo {year} {2014})}\BibitemShut {NoStop}%
\bibitem [{\citenamefont {Kurashige}\ \emph {et~al.}(2014)\citenamefont
  {Kurashige}, \citenamefont {Saitow}, \citenamefont {Chalupsky},\ and\
  \citenamefont {Yanai}}]{C3CP55225J}%
  \BibitemOpen
  \bibfield  {author} {\bibinfo {author} {\bibfnamefont {Y.}~\bibnamefont
  {Kurashige}}, \bibinfo {author} {\bibfnamefont {M.}~\bibnamefont {Saitow}},
  \bibinfo {author} {\bibfnamefont {J.}~\bibnamefont {Chalupsky}}, \ and\
  \bibinfo {author} {\bibfnamefont {T.}~\bibnamefont {Yanai}},\ }\href
  {\doibase 10.1039/C3CP55225J} {\bibfield  {journal} {\bibinfo  {journal}
  {Phys. Chem. Chem. Phys.}\ }\textbf {\bibinfo {volume} {16}},\ \bibinfo
  {pages} {11988} (\bibinfo {year} {2014})}\BibitemShut {NoStop}%
\bibitem [{\citenamefont {Dickhoff}\ and\ \citenamefont {{Van
  Neck}}(2008)}]{Bible}%
  \BibitemOpen
  \bibfield  {author} {\bibinfo {author} {\bibfnamefont {W.~H.}\ \bibnamefont
  {Dickhoff}}\ and\ \bibinfo {author} {\bibfnamefont {D.}~\bibnamefont {{Van
  Neck}}},\ }\href@noop {} {\emph {\bibinfo {title} {Many-body Theory
  Exposed!}}},\ \bibinfo {edition} {2nd}\ ed.\ (\bibinfo  {publisher} {World
  Scientific},\ \bibinfo {year} {2008})\BibitemShut {NoStop}%
\bibitem [{Note2()}]{Note2}%
  \BibitemOpen
  \bibinfo {note} {Using natural orbitals, and ordering them according to the
  NOON, is not an optimal choice for DMRG. It is better to group corresponding
  bonding and antibonding orbitals. However, this procedure allows for an
  unmonitored optimization.}\BibitemShut {Stop}%
\bibitem [{\citenamefont {Saitow}, \citenamefont {Kurashige},\ and\
  \citenamefont {Yanai}(2013)}]{DMRG-MRCI}%
  \BibitemOpen
  \bibfield  {author} {\bibinfo {author} {\bibfnamefont {M.}~\bibnamefont
  {Saitow}}, \bibinfo {author} {\bibfnamefont {Y.}~\bibnamefont {Kurashige}}, \
  and\ \bibinfo {author} {\bibfnamefont {T.}~\bibnamefont {Yanai}},\ }\href
  {\doibase 10.1063/1.4816627} {\bibfield  {journal} {\bibinfo  {journal} {J.
  Chem. Phys.}\ }\textbf {\bibinfo {volume} {139}},\ \bibinfo {pages} {044118}
  (\bibinfo {year} {2013})}\BibitemShut {NoStop}%
\bibitem [{\citenamefont {Bogaerts}\ \emph {et~al.}(2013)\citenamefont
  {Bogaerts}, \citenamefont {Van Yperen-De~Deyne}, \citenamefont {Liu},
  \citenamefont {Lynen}, \citenamefont {Van~Speybroeck},\ and\ \citenamefont
  {Van Der~Voort}}]{C3CC44473B}%
  \BibitemOpen
  \bibfield  {author} {\bibinfo {author} {\bibfnamefont {T.}~\bibnamefont
  {Bogaerts}}, \bibinfo {author} {\bibfnamefont {A.}~\bibnamefont {Van
  Yperen-De~Deyne}}, \bibinfo {author} {\bibfnamefont {Y.-Y.}\ \bibnamefont
  {Liu}}, \bibinfo {author} {\bibfnamefont {F.}~\bibnamefont {Lynen}}, \bibinfo
  {author} {\bibfnamefont {V.}~\bibnamefont {Van~Speybroeck}}, \ and\ \bibinfo
  {author} {\bibfnamefont {P.}~\bibnamefont {Van Der~Voort}},\ }\href {\doibase
  10.1039/C3CC44473B} {\bibfield  {journal} {\bibinfo  {journal} {Chem.
  Commun.}\ }\textbf {\bibinfo {volume} {49}},\ \bibinfo {pages} {8021}
  (\bibinfo {year} {2013})}\BibitemShut {NoStop}%
\end{thebibliography}%

\end{document}